\documentclass[twocolumn,pre,aps,floatfix,superscriptaddress,longbibliography]{revtex4-2}
\pdfoutput=1 %This is for the ArXiv submission only
\usepackage{amsmath,amssymb,eucal,graphicx,float,epstopdf}
\usepackage{epsfig,algorithm,algcompatible}
\usepackage[utf8]{inputenc}
\usepackage{siunitx}

\usepackage[colorlinks=true, urlcolor=blue, anchorcolor=blue, citecolor=blue,filecolor=blue,linkcolor=blue,menucolor=blue]{hyperref}

\usepackage{subfigure}

\def\beq{\begin{eqnarray}}
\def\eeq{\end{eqnarray}}
\def\be{\begin{equation}}
\def\ee{\end{equation}}
\def\bel{\begin{equation} \label}
\def\beel{\begin{eqnarray} \label}

\def\bm{\begin{math}}
\def\me{\end{math}}

\begin{document}

\title{ Stochastic Dynamics of Diffusive Memristor Blocks for Neuromorphic Computing}

\author{Wendy Otieno}
\affiliation{Department of Physics, Loughborough University, Loughborough LE11 3TU, UK}
\author{Alex Gabbitas}
\affiliation{Department of Physics, Loughborough University, Loughborough LE11 3TU, UK}
\author{Debi Pattnaik}
\author{Pavel Borisov}
\affiliation{Department of Physics, Loughborough University, Loughborough LE11 3TU, UK}
\author{Sergey Savel'ev}
\affiliation{Department of Physics, Loughborough University, Loughborough LE11 3TU, UK}
\author{Alexander G. Balanov}
\affiliation{Department of Physics, Loughborough University, Loughborough LE11 3TU, UK}
\begin{abstract}
Biological systems use neural circuits to integrate input information and produce outputs. Synaptic convergence, where multiple neurons converge their inputs onto a single downstream neuron, is common in natural neural circuits. However, understanding specific computations performed by such neural blocks and implementating them in hardware requires further research. This work focuses on synaptic convergence in a simplified circuit of three spiking artificial neurons based on diffusive memristors. Numerical modelling and experiments reveal input voltage combinations that enable targeted activation of spiking for specific neuron  configurations. We analyse the statistical characteristics of spiking patterns and interpret them from a computational perspective. The
numerical simulations match experimental measurements. Our findings contribute to development of universal functional blocks for neuromorphic systems.
\end{abstract}
\maketitle

\section{Introduction}

The emergence of artificial intelligence (AI) with deep learning (DL) and the Internet of Things (IoT) has led to the recent quest to develop high energy-efficient AI hardware capable to  process large-scale data (i.e., images, audio, and videos) in real time \cite{jo2024hardware,zhong2022spike,zhang2020artificial}. The main idea here is to deviate from the von Neumann architecture of computers by utilizing key intelligent components that integrate both memory and processing units \cite{sun2021future,seok2024beyond, chekol2022effect}.

Examples include the creation of artificial afferent nerves using NbO\textsubscript{x} Mott memristors \cite{zhang2020artificial}, the use of memristors for random number generator for IoT \cite{jiang2017novel}  or the use of memristors to perform the neural function of all or nothing \cite{pickett2013scalable}. Other examples are multi-input memristor-based artificial neurons for realising time-related perceptions such as directional selectivity and sound localization using volatile RRAMs \cite{wang2021neuromorphic}, NbO\textsubscript{x}-based oscillating neurons \cite{zhong2022spike,midya2025artificial}, or controlling neural connections via synaptic weights using 2D-hBN films \cite{jo2024hardware}. Additionally, electric circuits composed of solid-state neurons have been developed to mimic the property of biocircuit disease repair using biomedical implants that can adapt to biofeedback \cite{abu2019optimal}.

Recently, diffusive memristors have attracted significant attention as key components for intelligent systems, particularly in neuromorphic circuits \cite{chekol2022strategies}. This growing interest is driven by several advantageous properties of diffusive memristors, including their scalability, energy-efficient switching behavior \cite{chekol2022strategies}, compact footprint \cite{teja2024ultra, park2022experimental}, and high switching speed \cite{teja2024ultra}.

Importantly, these devices can mimick the behavior of biological synapses \cite{yang2019memristive, zins2023neuromorphic} and are capable of enabling more efficient data processing architectures \cite{yang2020neuromorphic, jeong2016memristors, seok2024beyond,midya2025artificial} by mitigating the von Neumann bottleneck \cite{sun2021future, seok2024beyond, chekol2022effect}.
Their ability to reproduce complex and biologically relevant behaviors--such as synaptic plasticity, spike-timing-dependent plasticity, frequency-dependent plasticity, and conductance relaxation \cite{ioannou2020evidence}--makes diffusive memristors highly suitable for neuromorphic computing applications \cite{Roadmap2024}.
These neuromorphic systems offer numerous potential advantages: power efficiency, adaptability, real-time processing (e.g., in autonomous driving or robotics), parallel processing \cite{akther2025modeling} (i.e., multitasking), high-speed computation, and fault tolerance. However, realizing brain-inspired computing devices requires research in designing a specific neuromorphic architecture composed of diffusive memristors' blocks, arranged in a configuration that would enable neural-like information processing.

Typical diffusive memristors  are two-terminal structures that operate based on the diffusion of ions or atomic species within the material sandwiched between two contacts (usually Pt or Au) \cite{wang2017memristors}. Unlike traditional memristors, which often have non-volatile memory (they retain their resistance state after power is off), diffusive memristors typically exhibit volatile switching. That means they return to their original state after a short time, making them useful for mimicking biological synapses. These devices usually consist of a metal filament (e.g., silver or copper) embedded in a solid electrolyte (e.g., Ag:SiO$_2$ or Cu:SiO$_2$) or insulating matrix like SiO$_2$, HfO$_2$. When a voltage is applied, metal atoms diffuse through the dielectric, forming a temporary conductive path (filament). This path dissolves once the voltage is removed due to diffusion, making the low-resistance effect transient. Because of intricate interplay of nano-mechanical, heat and electric degrees of freedom \cite{akther2021deterministic,savel2013mesoscopic} the diffusive memristors exhibit various spiking regimes (including noise-mediated behaviour) typical  for biological neurons with  dynamical properties, which  have been shown to be equivalent to Ca$^{2+}$ functionality in biological synapses \cite{wang2017memristors}.

\begin{figure*}[t!]
    \centering
    \includegraphics[width=6.75cm]{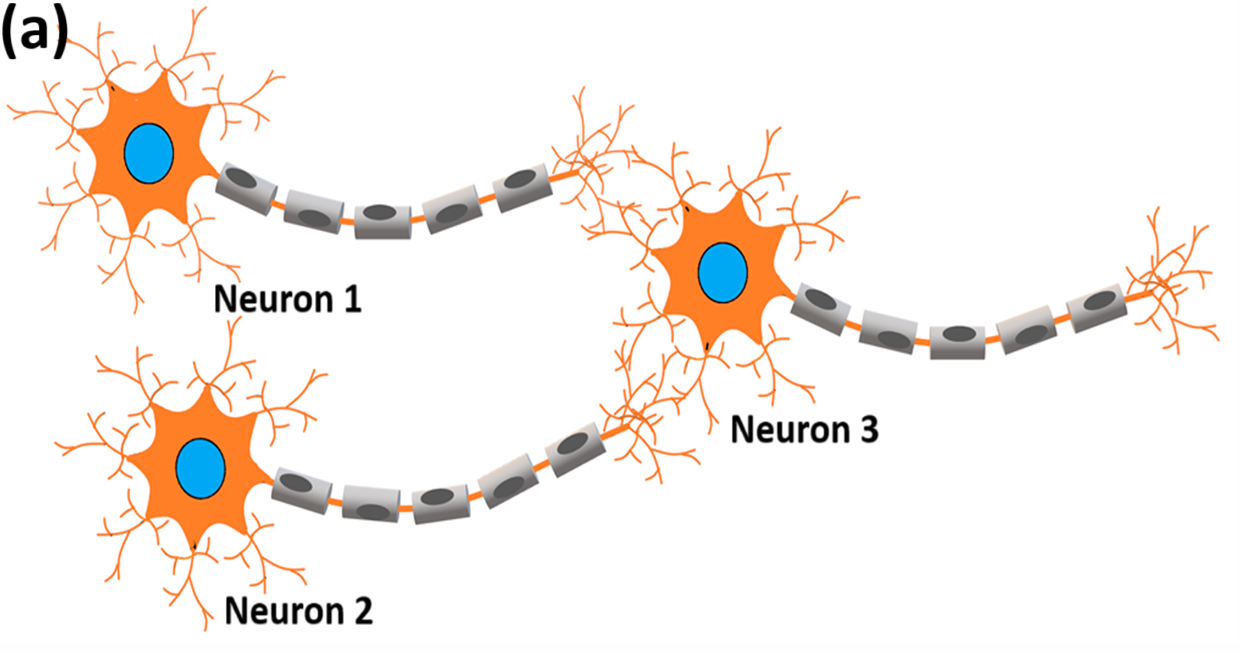}
    \includegraphics[width=6.75cm]{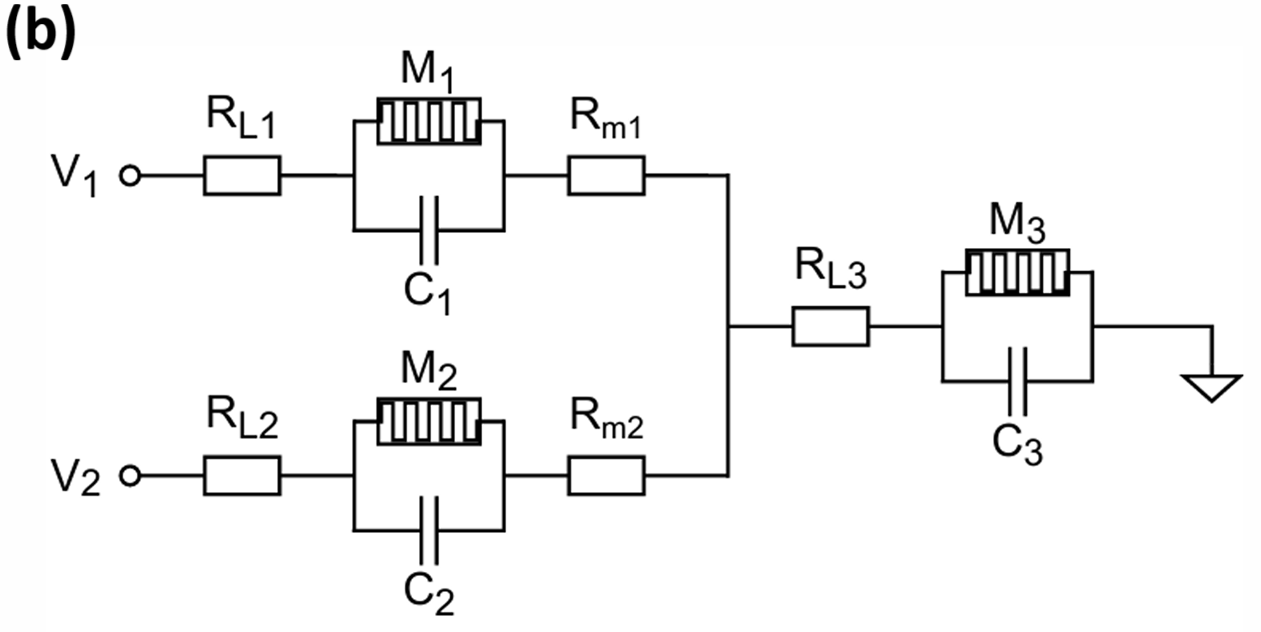}
    \includegraphics[width=4.15cm]{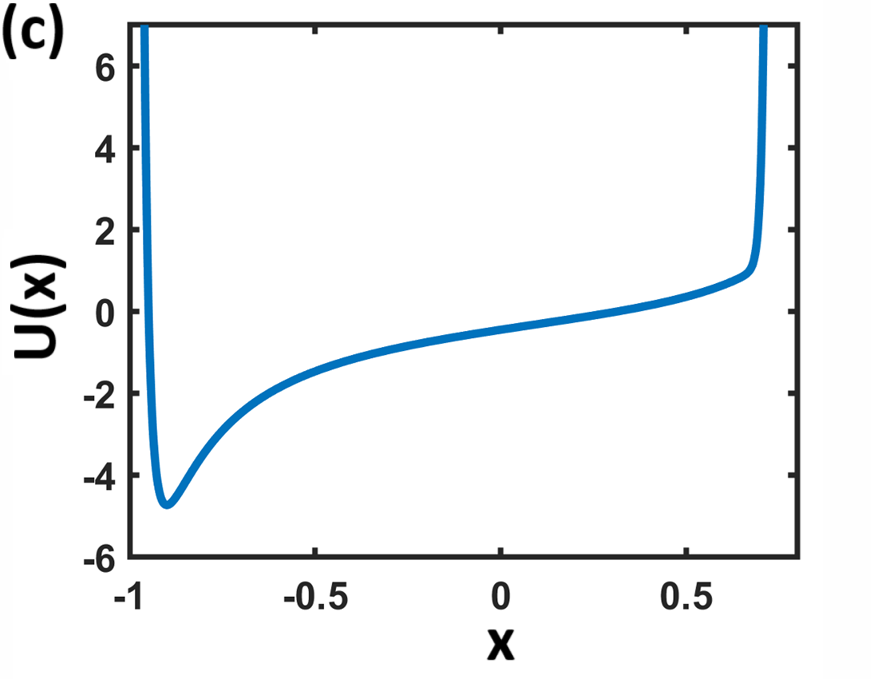}
    \caption{ (a) Synaptic convergence involving three neurons. (b) The neuromorphic block modeling the synaptic convergence: each artificial neuron consists of a diffusive memristor $R_{M_{i}}$ in series to a load resistor $R_{i}$ and in parallel to a capacitor $C_{i}$ ($i = 1,2,3$). External input voltages $V_{1}$ and $V_{2}$ are applied to the neuromorphic circuits mimicking Neuron 1 and Neuron 2 shown in the middle panel, respectively. The artificial neuron involving the memristor $R_{M_{3}}$ correspond to the post-synaptic Neuron 3 depicted in the middle panel. (c) The electrochemical potential profile U(x) of the diffusive memristor used in simulation \cite{ushakov2021role}.}
    \label{Fig:neuromorphiccircuit}
\end{figure*}

Biological neurons are naturally noisy. Noise is a fundamental aspect of how neurons in networks function and communicate with one another.  For example, in the auditory system, noise through auditory gain control aids in focusing on relevant stimuli \cite{auerbach2022hearing} (i.e. listening to someone’s important message) while tuning out irrelevant stimuli (i.e. background noise from other conversations being held). Noise benefits networks of neurons by enabling probabilistic interference through sampling (neurons decision is hinged on the probability of different outcomes) \cite{maass2014noise}, self-organization and learning \cite{maass2014noise} and signal enhancement \cite{benzi1981mechanism,shu2003barrages,faisal2008noise}. 

The degree of freedom related to heat transfer and the probabilistic nature of conducting filament formation makes the dynamics of diffusive memristors inherently stochastic, similar to living neurons \cite{kim2021self,jiang2017novel}. Consequently, like living neurons, circuits of diffusive memristors should be capable of using random fluctuations for computational purposes. However, neuromorphic technologies still demand insights into the mechanisms enabling simple blocks of neurons to contribute to computing performance. 
  
Here, we explore the stochastic dynamics of a simple synaptic convergence involving three artificial neurons based on diffusive memristors, as illustrated in Fig.~\ref{Fig:neuromorphiccircuit}. This setup could, for example, model  a component of the sensory system, where two primary neurons, receiving sensory input from peripheral receptors, synapse onto a secondary post-synaptic neuron. Through numerical simulations and experiments, we investigate how firing and spiking in this three-neuron system are influenced by the voltages $V_1$ and $V_2$ applied to the primary neurons. We identify the voltage ranges that trigger the firing of either a combination or all neurons and uncover the potential to utilise these dynamic mechanisms for neuromorphic computing.

\section{Methods}

\subsection{Experimental Details}
\label{ss:exp}

\begin{figure}[t!]
    \centering
    \includegraphics[width=4cm]{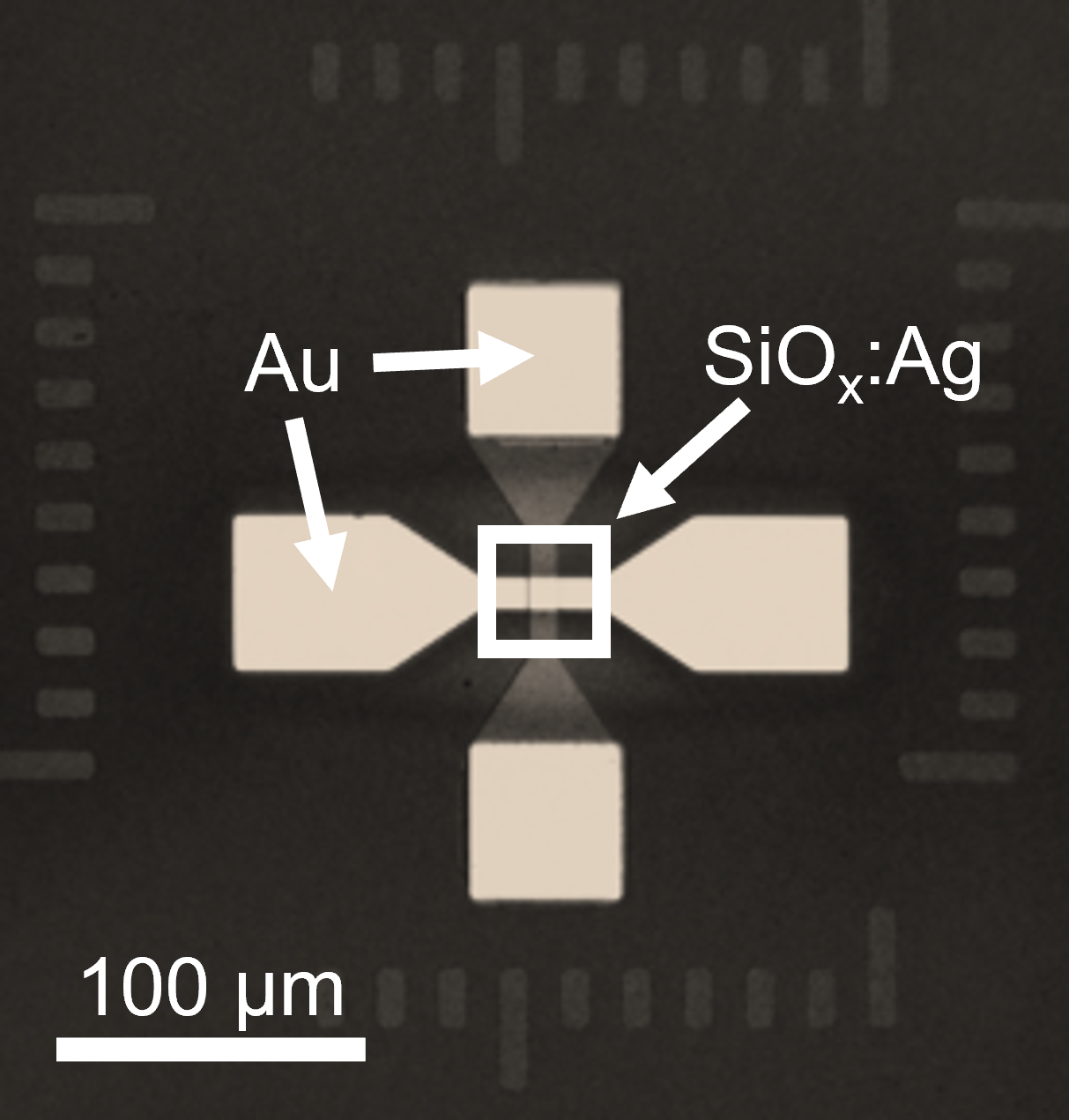}
    \caption{The fabrication of each diffusive memristor occurred via magnetron sputtering deposition and UV photolithography techniques. Bottom electrodes of 5 nm Ti / 45 nm Au  were deposited onto SiO\textsubscript{2}/Si wafers followed by a switching layer of 50 nm SiO\textsubscript{x}:Ag  before top electrodes of thicknesses 5 nm Ti / 120 nm Au.}
    \label{Fig:diffsample}
\end{figure}

In the experimental realization of our neural network, three identically-deposited diffusive memristors were connected in a configuration matching that of Fig~\ref{Fig:neuromorphiccircuit}. These memristors, denoted in the schematic circuit  as $R_{M_{1,2,3}}$, were made through a combination of magnetron sputtering deposition and UV photolithography techniques. Bottom electrodes of 5 nm Ti / 45 nm Au were patterned by photolithography and deposited onto SiO\textsubscript{2}/Si wafers, where the Ti layer was applied to improve adhesion to the substrate. These 200 $\mu$m long paddle-shaped electrodes possessed a width of 10 $\mu$m at their centres. Atop this, a switching layer of 50 nm SiO\textsubscript{x}:Ag was deposited by co-sputtering SiO\textsubscript{2} and Ag with deposition rates of 0.6 A/s and 0.1 A/s, respectively. Finally, top electrodes of the same geometry as the bottom, yet thicknesses of 5 nm Ti / 120 nm Au, were deposited perpendicularly to the bottom electrodes. Such a process created a crossbar junction of 10 $\mu$m x 10 $\mu$m SiO\textsubscript{x}:Ag between the patterned electrodes (see Fig~\ref{Fig:diffsample}).

Each device was connected in parallel to a capacitor \textit{C\textsubscript{1,2,3}} = 1 nF and in series to a load resistor $R_{1,2,3}$ (see Fig~\ref{Fig:neuromorphiccircuit}) in order to create artificial neurons which represent Pearson-Anson oscillators \cite{liu2016threshold}, promoting spiking behaviour within each memristor. These diffusive memristors possess internal capacitances of approximately 24 pF when applied voltage $\leq$ 1.0 V \cite{gabbitas2023resistive}. Therefore, the parallel capacitors attached are significantly larger than the internal capacitance of the devices, effectively negating this parasitic value.

\begin{figure}[t!]
    \centering
    \includegraphics[width=9.5cm]{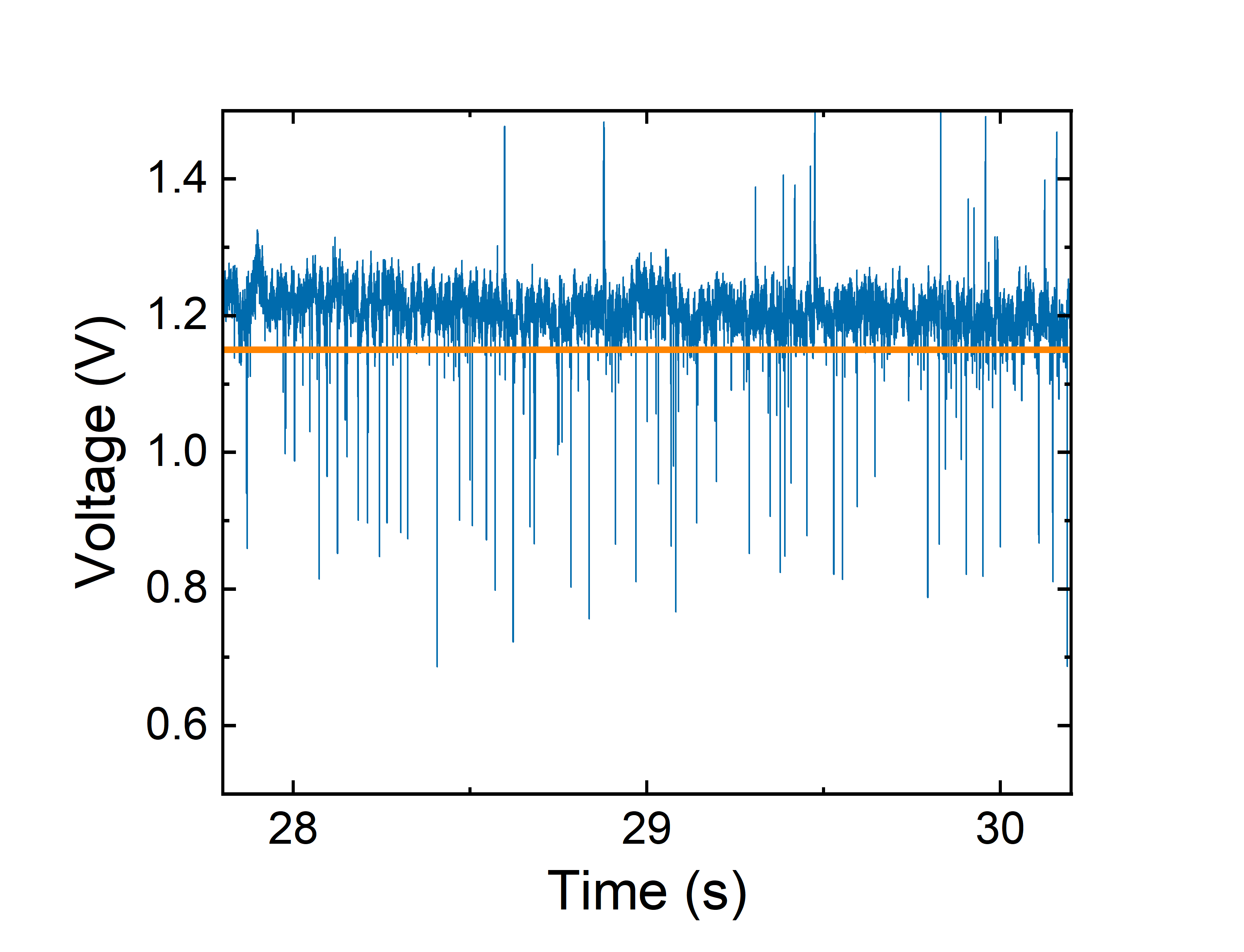}
    \caption{Experimentally measured voltage oscillations for memristive device $M_{3}$ at voltages $V_{1}$ = 1.2 V and $V_{2}$ = 3.0 V. To receive the classification of `spiking', the amplitude of oscillations in two or more of the ON periods must be $>$ 0 V and at least $10\%$ higher than the baseline noise level observed during OFF periods (indicated by orange threshold line). In addition to this single ON period example, the device meets this criteria in each other ON period, and is thus deemed to be spiking.}  
    %\caption{ {\color{blue} Experimentally measured voltage oscillations for each memristive device ($M_{1}, M_{2}, M_{3}$, in (a), (b) and (c), respectively) at voltages $V_{1}$ = 0.2 V and $V_{2}$ = 2.4 V. To receive the classification of `spiking', the amplitude of oscillations in two or more of the on periods must be $>$ 0 V and at least $10\%$ higher than the baseline noise level observed during off periods. In this example, both $R_{M_{1}}$ and $R_{M_{2}}$ do not demonstrate such conditions, therefore are not deemed to be consistently spiking. However the third device ($M_{3}$) meets this criteria in each on period and is thus deemed to be the only memristor spiking at these voltages.}}
    \label{Fig:spikrealiz}
\end{figure}

Long voltage pulses of duration \textit{t\textsubscript{ON}} = 10 s were applied simultaneously to both series load resistors $R_{1}$ and $R_{2}$ in order to initiate self-oscillating (i.e. neuron spiking) behaviour in both memristors $R_{M_{1}}$ and $R_{M_{2}}$. Meanwhile, $R_{M_{3}}$ was solely supplied by the outputs of the two previous memristors. Three sequential pulses were applied for a given voltage, each separated by an `OFF' period of \textit{t\textsubscript{OFF}} = 10 s, in order to increase statistics and lessen the effects of stochasticity. The load resistors used were of magnitudes \textit{R\textsubscript{1}} = \textit{R\textsubscript{3}} = 55 k$\Omega$ and \textit{R\textsubscript{2}} = 60 k$\Omega$. Experimentally, the amplitudes of voltage pulses \textit{V\textsubscript{1}} and \textit{V\textsubscript{2}} were varied from 0 V to 3.6 V with a step size of 0.2 V, in order to create a wide range of input combinations. For all combinations of input voltages, the presence of spiking was investigated for the three memristors, where each device received a classification of `spiking' or `not spiking' after analysing their voltage outputs. 

To classify spiking behaviour, we used criteria that voltage oscillations for a particular device were required to be at least 10\% higher than the baseline level of noise observed during OFF periods. Such a condition must be met in at least two of the three ON periods for a given voltage combination to be classified as spiking. An example of a voltage measurement for device $M_{3}$ is given in Fig. \ref{Fig:spikrealiz}, for a single combination of input voltages. In Fig. \ref{Fig:spikrealiz}, the memristor $M_{3}$ displays spiking behaviour of sufficient amplitude, indicated by consistently crossing the orange threshold line. In this particular example, such behaviour is replicated in all other ON periods, and is thus deemed to be spiking consistently at the given input voltages.

%Examples of voltage measurements for each device are given in Fig. \ref{Fig:spikrealiz}, for a single combination of input voltages. In Fig \ref{Fig:spikrealiz}(a,b), the spiking classification conditions are not met, therefore $M_{1}$ and $M_{2}$ are judged not to be spiking. Meanwhile, the third memristor does display spiking behaviour of sufficient amplitude, in the positive polarity, for all ON periods (Fig. \ref{Fig:spikrealiz}(c)) and is thus deemed to be spiking consistently. This example voltage combination gives the overall result of only $M_{3}$ spiking, which is later compared to all over applied combinations in order to map the behaviour of all three devices.

Spiking in both $R_{M_{1}}$ and $R_{M_{2}}$ was recorded using current measurement with Keithley 7410 digital multimeters (DMMs) across 1 k$\Omega$ monitoring resistors \textit{R\textsubscript{a}} and \textit{R\textsubscript{b}}, respectively (see Fig.~\ref{Fig:neuromorphiccircuit}). In contrast, the prospective spiking in $R_{M_{3}}$ was measured directly across the memristor itself, via a PicoScope 5443D.

\subsection{Model Details}
\label{ssec:model}
In our numerical simulations, we employed the charge transport model \cite{savel2013mesoscopic}, which has previously been successful in reproducing and explaining experimental results involving diffusive memristors \cite{wang2017memristors, akther2021deterministic, pattnaik2023temperature}. In this framework, each diffusive memristor is represented by a set of stochastic differential equations that describe the motion of a metallic cluster between a nearly formed conductive filament and a contact \cite{yi2016quantized}. The particle moves in an electrochemical potential formed between the filament tip and a contact, being driven by the applied electric field. Charge transport occurs via electron tunneling through the particle. For the circuit presented in the left panel of Fig.~\ref{Fig:neuromorphiccircuit} the dimensionless model equations read 

\begin{align}
\frac{dx_{i}}{dt} &= - \beta \frac{dU}{dx_{i}} + \frac{q_{i} V_{M_{i}}}{L} + \sqrt{2 k_{B} \eta_{i} T_{i}} \xi_{i}(t)  \label{xirealization},  \\
\frac{dT_{i}}{dt} &= \frac{V_{M_{i}}^2}{C_{h}R_{M_{i}}(x_{i})} - k(T_{i}-T_{0})  \hspace{0.5cm} (i = 1, 2, 3) \label{Tirealization},  \allowdisplaybreaks  \\
 \frac{dV_{M_{l}}}{dt} &= \frac{b_{l}}{\tau_{4}} - \frac{a_{l}V_{M_{l}}}{\tau_{4}} + \sum_{j = 1, j \neq l}^{2} \bigg(\frac{2\overline{R_{3}}C_{j}V_{M_{j}}}{\tau_{4}}  - \frac{2V_{M_{3}}\tau_{j}}{\tau_{4}} \bigg) \nonumber  \\ 
 & \hspace{4.8cm} (l = 1,2) \label{Vm1realization}, \\
\frac{dV_{M_{3}}}{dt} &= \sum_{j=1}^2 V_{M_{j}}\overline{R_{3}}\bigg(\frac{1}{\tau_{3}R_{M_{j}}} - \frac{C_{j} a_{j}}{\tau_{3}\tau_{4}} + \frac{2C_{1}C_{2}\overline{R_{3}}}{\tau_{3} \tau_{4}} \bigg)  \nonumber \\ & \hspace{0.5cm}  + \sum_{j=1}^2 \frac{C_{j}b_{j}\overline{R_{3}}}{\tau_{3}\tau_{4}}   - V_{M_{3}}\overline{R_{3}}d_{3} \label{Vm3realization},
\end{align}
where  $x_i$, $V_{M_{i}}$ and $T_i$ is the position of the mobile particle, the voltage drop and the temperature in the $i^{th}$ memristor respectively.

The particles are subjected to an effective electrical force $q_iV_{M_{i}}/L$ with induced effective particle charge $q_i$ and $L$ being the gap between the tip of the filament and the contact. The particle's motion is affected by random fluctuations modelled by the delta-correlated Gaussian white noise $\xi_{i}(t)$ with average $\langle \xi_{i}(t) \rangle = 0$ and $ \langle \xi_{i}(t)\xi_{j}(t + t') \rangle = \delta(t-t')\delta_{ij}$, whose intensity depends on the temperature $T_i$ and a viscosity parameter $\eta_i$, while $k_{B}$ denotes the Boltzmann constant, and $\langle.\rangle$ designates an ensemble average. 

The function $U(x)$ accommodates a phenomenological electrochemical potential formed within the gap between the tip of the conducting filament and the contact \cite{ushakov2021role} 
 \begin{equation}
U(x) = \frac{0.05}{(x+1)^2} - \frac{1}{(x+1)} - \frac{0.5}{(x-1)} + (1.2x+0.168)^{100}. \label{UPotential}
 \end{equation}

It has a minimum near the tip of the filament, which for low voltages $V_{M_{i}}$ attributes to a stable equilibrium between the attracting chemical (ionic) force within the filament and the repelling Coulomb force. For sufficiently large voltages $V_{M_{i}}$, an additional potential gradient $q_iV_{M_{i}}/L$ causes the particle to leave the filament’s tip and drift towards the contact. However, if $V_{M_{i}}$ decreases or changes polarity, the particle is pushed back from the contact to the filament. 
  
To reflect the random character of the filament formation and evolution \cite{savel2011molecular}, which affects the drift-diffusion of the particle, we introduce a random parameter $\beta$. This parameter reflects the reconfiguration of nano-clusters near the filament tip, which, for example, alters the shape of the particle attachment region. This, in turn, inevitably changes the potential associated with minimising the surface energy that drives the particle’s attraction to the filament tip \cite{kim2021self}. A factor of $\beta$ is multiplied to the derivative of the potential $U'(x_{i})$ where $\beta \in [0.05,1]$ is a uniformly generated random number. In our calculation, the randomization of the potential occurs at arbitrary times $t = 0.1$ - $0.2$, $t = 0.3$ - $0.4$, $t = 0.5$ - $0.6$, ... while $\beta = 1$ otherwise. 

The temperature dynamics defined by \eqref{Tirealization} is governed by the Newton’s cooling law, where the rate of heat transfer to the sink is determined by the cooling constant $k$ and the background temperature $T_0$.  With this, the heat source is governed by the Joule dissipation and is coupled to temperature via the parameter of thermal capacitance $C_{th}$. The function $R_{M_{i}}(x_{i})=R_0\cosh(x_i/ \lambda$) characterizes tunneling resistance of $i^{th}$  memristor with $R_0$ being the maximal tunneling resistance and $\lambda$ -- tunneling length. 

The equations \eqref{Vm1realization} - \eqref{Vm3realization}, describing the dynamics of the voltage drop $V_{M_{i}}$ across the memristors, are derived from the Kirchhoff’s circuit rules for the schematic in Fig.~\ref{Fig:neuromorphiccircuit}, which come with the following parameters expressed in terms of circuit elements: 

{\allowdisplaybreaks
\begin{align*}
\overline{R_{1}} &= (R_{1}+R_{a}), \overline{R_{2}} = (R_{2}+R_{b}), \overline{R_{3}} = (R_{3}+R_{c}) \\
\tau_{1} &= \overline{R_{1}}C_{1}, \tau_{2} = \overline{R_{2}}C_{2}, \tau_{3} = \overline{R_{3}}C_{3}  \\
\tau_{4} &= 2 C_{1}C_{2} (\overline{R_{1}}\overline{R_{2}}+\overline{R_{1}}\overline{R_{3}}+\overline{R_{2}}\overline{R_{3}}) \\
b_{1} &= 2V_{1}\tau_{2} + 2V_{1}C_{2}\overline{R_{3}}- 2C_{2}\overline{R_{3}}V_{2} \\
b_{2} &= 2V_{2}\tau_{1} + 2V_{2}C_{1}\overline{R_{3}} - 2C_{1}\overline{R_{3}}V_{1} \\
a_{1} &= 2\tau_{2} + \frac{2 \tau_{2} \overline{R_{1}}}{R_{M_{1}}} + \frac{2\overline{R_{3}}\tau_{2}}{R_{M_{1}}} + \frac{2C_{2}\overline{R_{1}}\overline{R_{3}}}{R_{M_{1}}} + 2C_{2}\overline{R_{3}} \\
a_{2} &= 2\tau_{1} + \frac{2 \tau_{1} \overline{R_{2}}}{R_{M_{2}}} + \frac{2\overline{R_{3}}\tau_{1}}{R_{M_{2}}} + \frac{2C_{1}\overline{R_{2}}\overline{R_{3}}}{R_{M_{2}}} + 2C_{1}\overline{R_{3}} \\
d_{3} &= \frac{1}{\tau_{3} R_{M_{3}}} + \frac{2 \tau_{2} C_{1}}{\tau_{3} \tau_{4}} + \frac{2C_{2}\tau_{1}}{\tau_{3} \tau_{4}}
\end{align*}
}

For numerical modeling  we use the following set of dimensionless parameters:
$q_{i}/L = 0.2$, $k = 0.2$, $C_{h} = 0.18$, $T_{0} = 1.1$, $\tau_{1} = \tau_{2} = \tau_{3} = 1$, $\lambda = 0.13$ and $R_0=1$ with initial conditions: $x_{i}(0) = -0.95$, $T_{i}(0) = 30$ and  $V_{M_{i}}(0) = 3.1$ ($ i = 1\ldots3$) and resistances: $\overline{R}_{1} = 560$, $\overline{R}_{2} = 610$, $\overline{R}_{3} = 560$. This parameter selection assumes that the neurons are non-identical, with slight differences in their spiking thresholds and timescales. Under these conditions, we study the spiking behavior of our three-neuron circuit by varying the external voltages $V_{j}$  $(j = 1,2)$ and $2k_{B}\eta_{i}$  ($i = 1,2,3$). The stochastic model equations has been integrated numerically  using the Euler-Maruyama method implementing It\^o calculus, with a time step $h=0.001$.

To compare the regularity and variability of the spiking regimes we introduced two characteristics, which have previously been used to evaluate stochastic properties of spiking activity of living neurons \cite{Maimon:2009aa,lengler2017note} and in comparative analysis of living and artificial neurons \cite{midya2025artificial}. The first characteristic is the ratio of the standard deviation of inter-spike intervals (ISI) to the mean ISI
\begin{equation}
CV_{1} = \frac{(\overline{\Delta t}^2 - (\overline{\Delta t})^2)^{1/2} }  {\overline{\Delta t}},
\label{eq:cv1}
\end{equation}
 where $(\overline{\Delta t}^2 - (\overline{\Delta t})^2)$ is the variance of the interspike intervals, $\overline{\Delta t}$ is the expected value of the interspike intervals. This metric quantifies the degree of non-periodicity of the entire spiking realization, and also evidences a deviation from Poisson statistics. The second characteristic describes the local variability of ISIs and is defined as 
\begin{equation}
CV_{2} = \bigg\langle 2 \frac{|\Delta t_{i} - \Delta t_{i-1}|}{\Delta t_{i} + \Delta t_{i-1}}  \bigg\rangle,
\label{eq:cv2}
\end{equation}
where  $\Delta t_{i-1}$ and $\Delta t_{i}$ are consecutive interspike intervals. The quantity $CV_2$ describes the local correlations in sequential ISIs and reflects the persistence of ISIs changes. 

We also compute a probability density estimate of our bivariate data ($CV_{1}$ and $CV_{2}$) using the ``ksdensity" function on Matlab R2022b to analyse the correlation between $CV_{1}$ and $CV_{2}$.

\begin{figure}
    \centering
    \includegraphics[width=7.85cm]{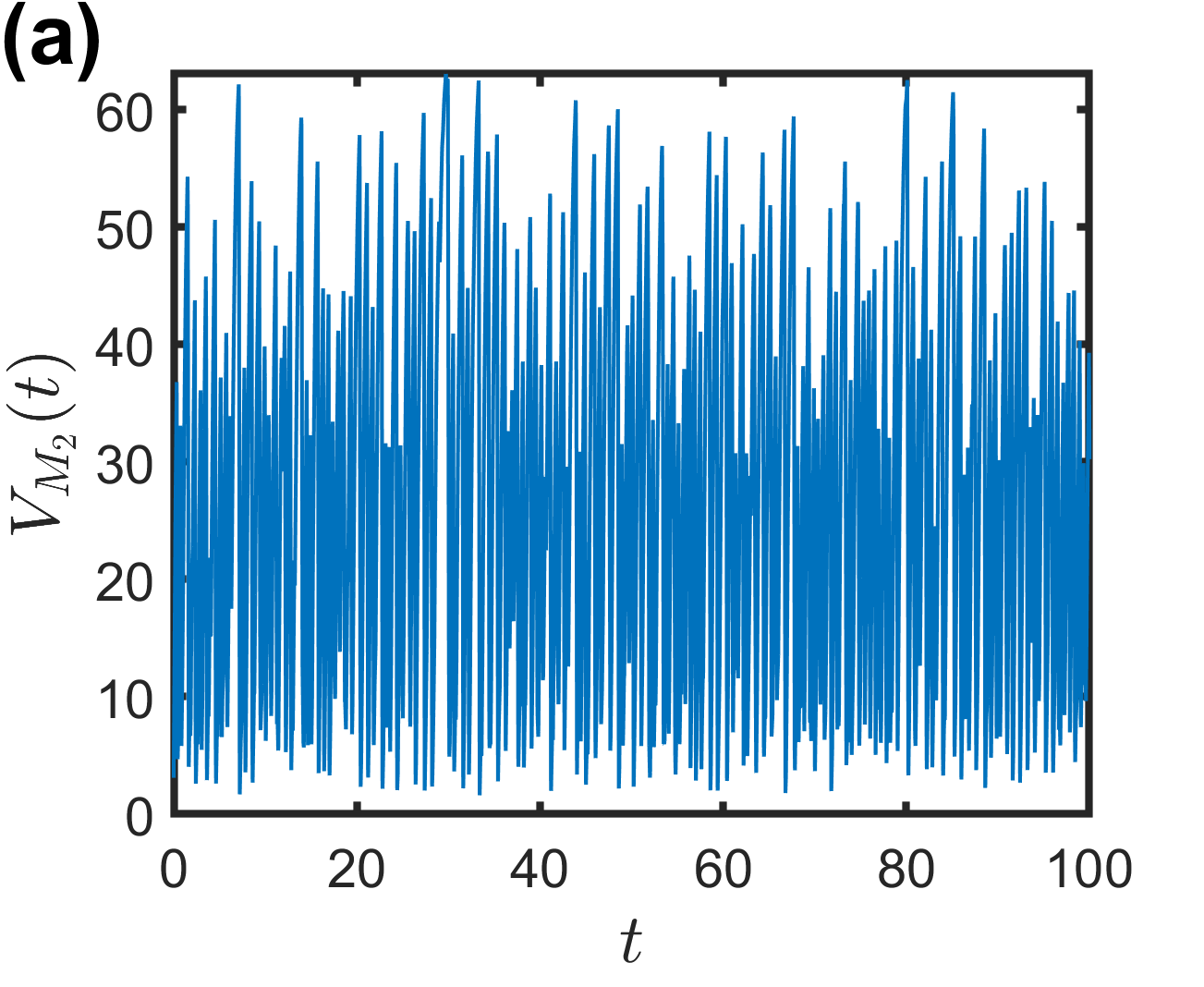}
    \includegraphics[width=7.85cm]{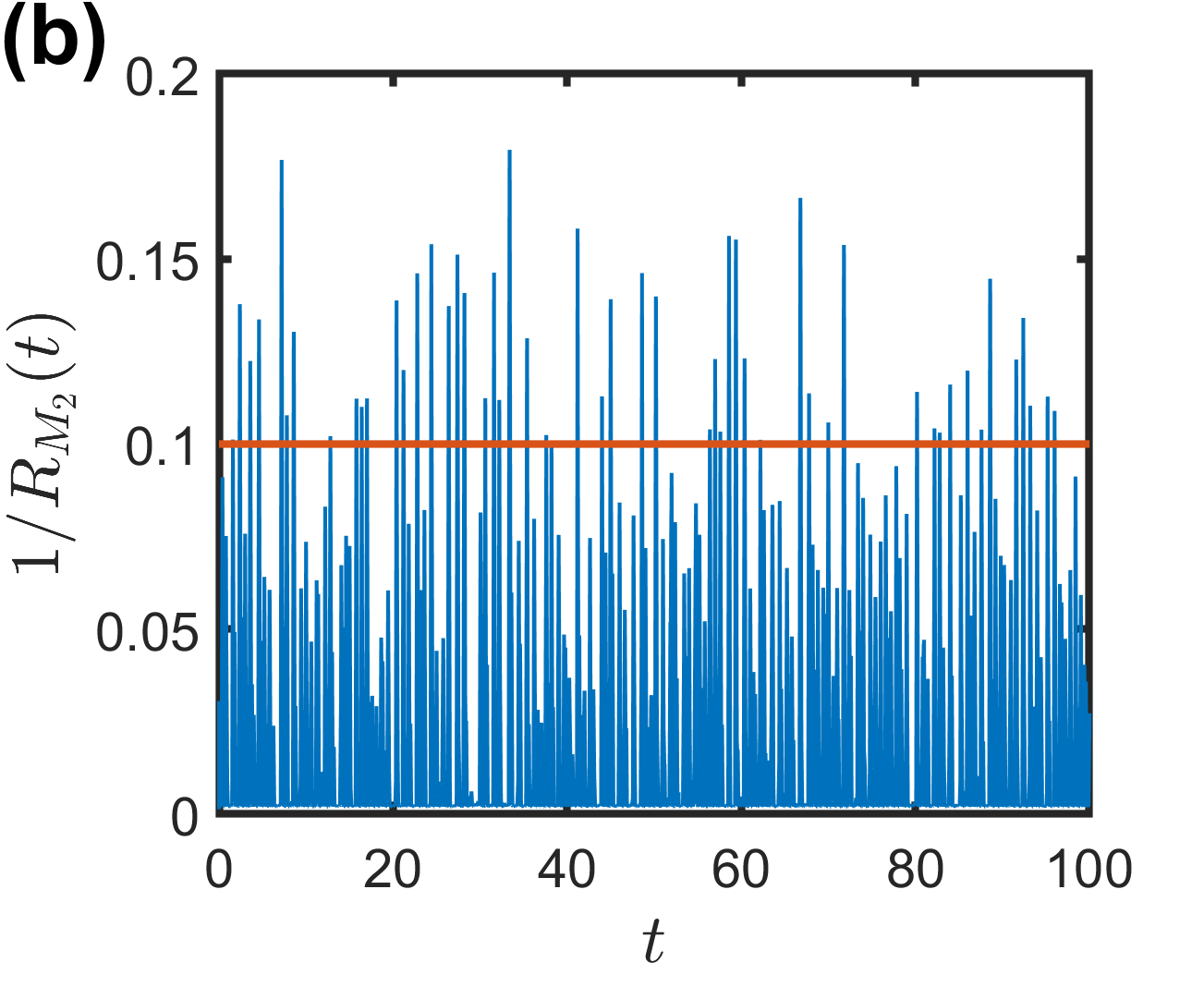}
    \caption{Spiking in the second memristor $V_{M_{2}}(t)$ (a) is realized in the conductance $1/R_{M_{2}}(t)$ (b) where the local maxima meets the spiking criteria by exceeding the threshold $0.1$ (given by the orange horizontal line). This example is observed for parameters $V_1= 100$, $V_2=300$ and $2k_{B}\eta_{i} = 10^{-5}$.}
\label{Fig:interspikeVM2RM2}
\end{figure}

 A typical spiking realization generated by the model \eqref{xirealization}-\eqref{Vm3realization} is presented in Fig.~\ref{Fig:interspikeVM2RM2}, which shows the time series of $V_{M_{2}}(t)$ (a) and $1/R_{M_{2}}(t)$ (b) calculated for $V_1= 100$ and $V_2=300$ with noise intensities $2k_{B}\eta_{i} = 10^{-5}$. Similar to the experiment, the spikes in each memristor were detected as local maxima occurring above the baseline level of noise. In numerical simulations, it was more convenient to detect these spikes in the realizations of conductance $1/R_{M_{i}}(t)$ as conductance produces immediate and measurable current changes that align with the computational behaviour of diffusive memristors. If the maximum of $1/R_{M_{i}}$ exceeds the threshold $0.1$, then we considered that the memristor produces  a spike. Such consideration correspond to  our experimental requirement for spiking where the amplitude of voltage oscillations should be at least 10\% higher than the baseline level of noise observed during the OFF periods.

\begin{figure*}
    \centering
    \includegraphics[width=5.85cm]{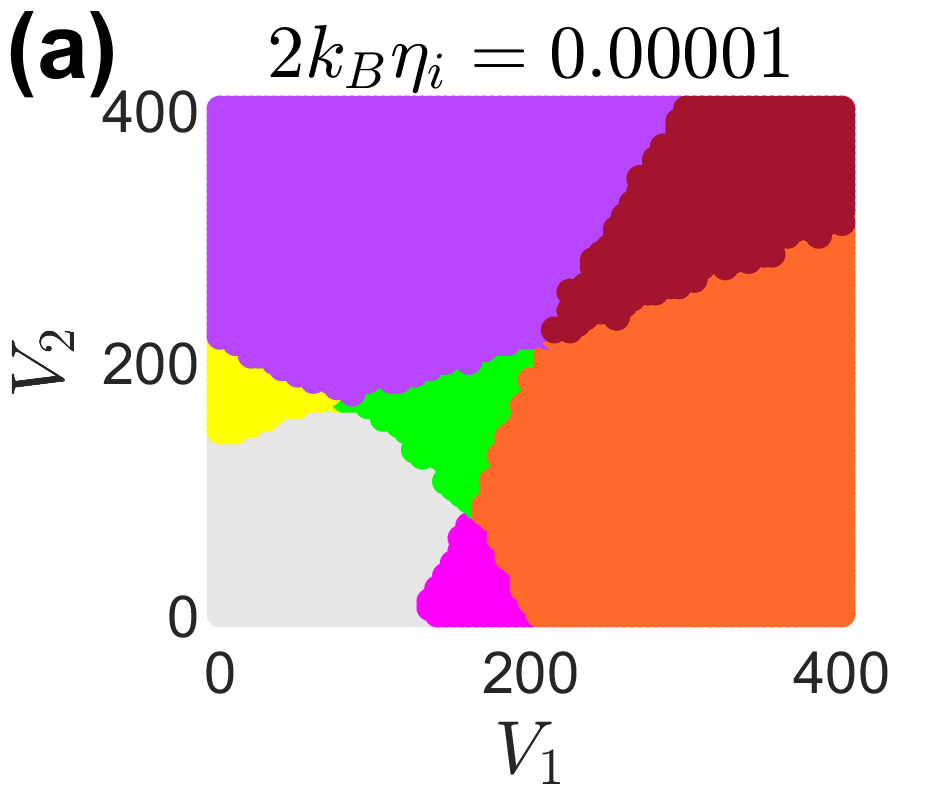}
    \includegraphics[width=5.85cm]{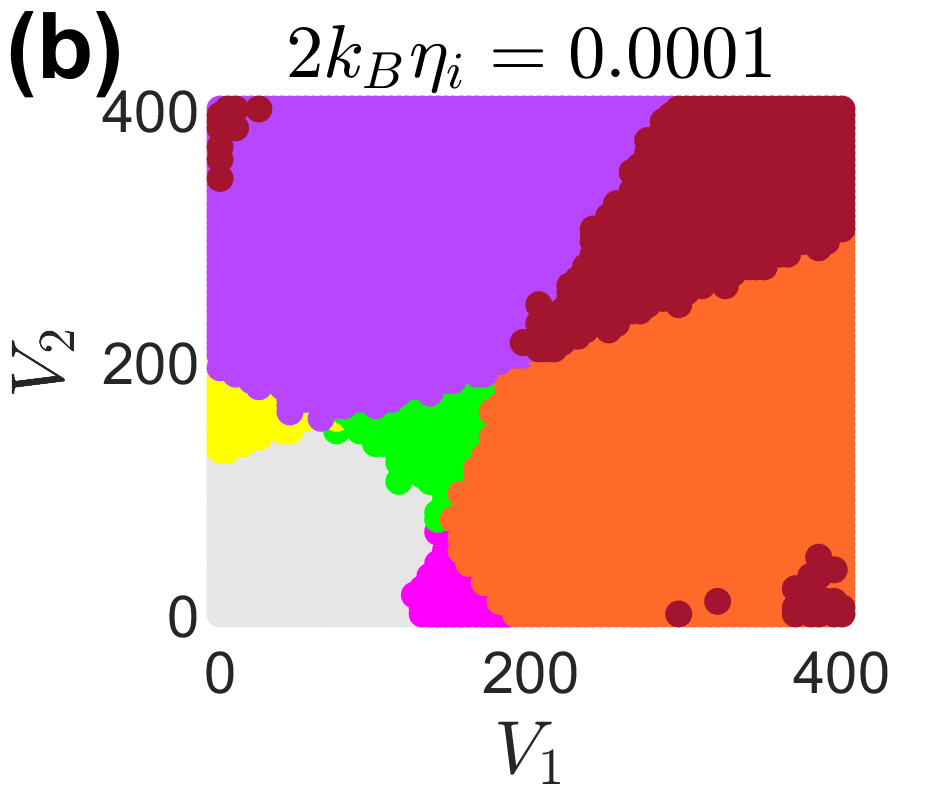}
    \includegraphics[width=5.85cm]{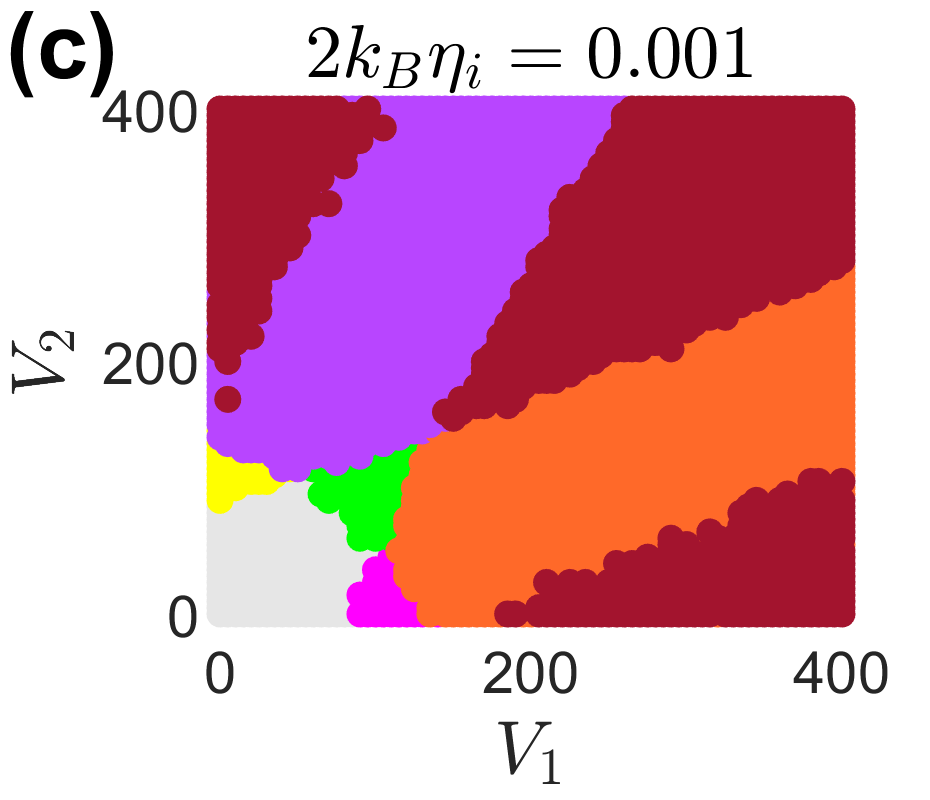}\\
    \includegraphics[width=12.85cm]{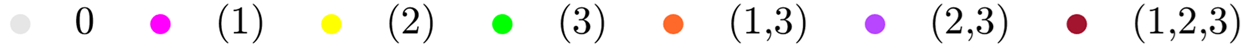}\\
    \vspace{0.25cm}
    \includegraphics[width=5.85cm]{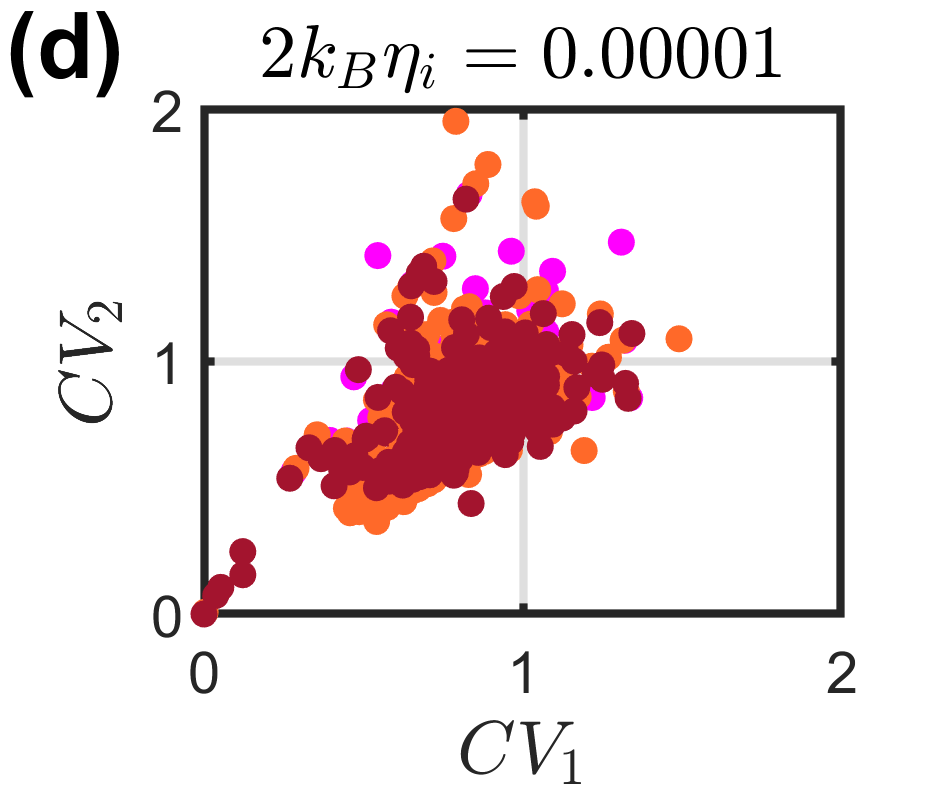}
    \includegraphics[width=5.85cm]{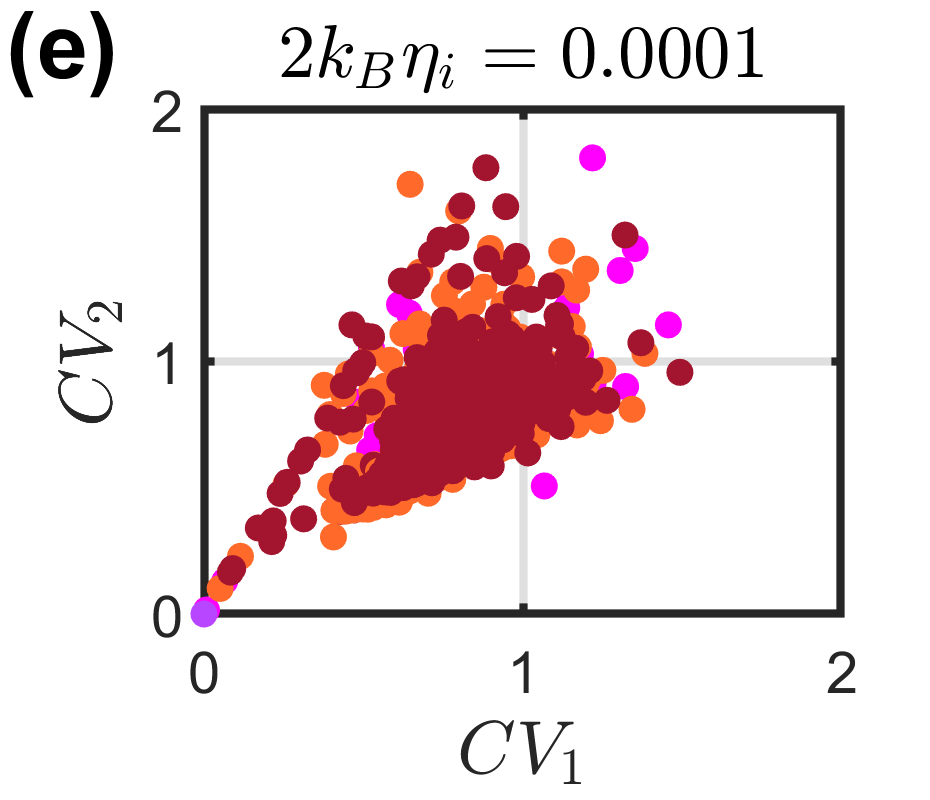}
    \includegraphics[width=5.85cm]{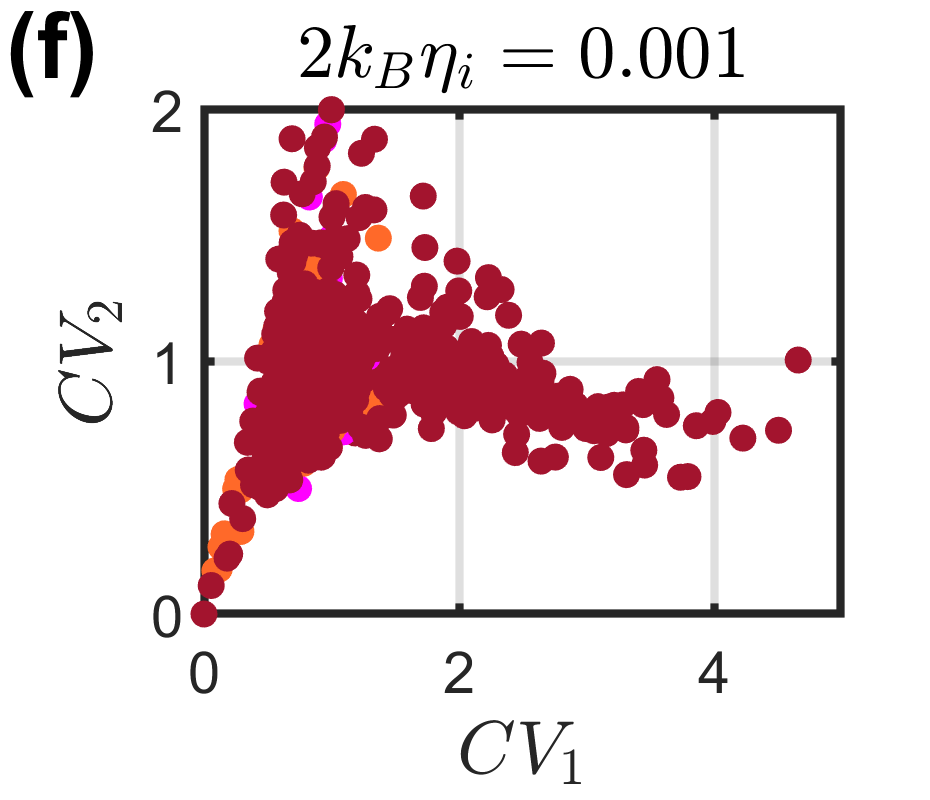}
    \caption{ Areas of existence for various spiking patterns on the parameter plane $(V_1,V_2)$, calculated numerically for (a) $2k_{B}\eta_{i} = 10^{-5}$, (b) $2k_{B}\eta_{i} = 10^{-4}$, and (c) $2k_{B}\eta_{i} = 10^{-3}$. Different patterns are indicated by different colors. The corresponding $(CV_{1},CV_{2})$ values for (d)-(f) show that as noise-induced spiking dominates, the values of $(CV_{1},CV_{2})$ increase. We show $(CV_{1},CV_{2})$ values for memristor $R_{M_1}$ when $(1)$, $(1,3)$ and $(1,2,3)$-spiking occur.
    }
    \label{Fig:shakenpotential}
\end{figure*}

\section{Results and Discussion}
\subsection{Numerical simulations.}
\label{ssec:numl}
To gain insight into the spiking patterns generated in the synaptic convergence modelled by the memristive block in Fig. \ref{Fig:neuromorphiccircuit}, we investigated the existence of spiking in each memristor over a wide range of voltage parameters $V_1$ and $V_2$.  

We started our analysis with  numerical simulation using  the model equation  \eqref{xirealization}-\eqref{Vm3realization}, where we set the same level of noise  $2k_{B}\eta_{1} = 2k_{B}\eta_{2} = 2k_{B}\eta_{3}$  for all memristors, and assume $\beta \in [0.05\;1]$. Fig.~\ref{Fig:shakenpotential} (a) summarizes the results of numerical studies  for $2k_{B}\eta_{i} = 10^{-5}$, which were performed in the voltage range $V_{1}, V_{2} \in [0,400]$ with voltage step 5. The spiking is registered by evaluation of the $1/R_{M_{i}}(t)$--realization, as described in subsection \ref{ssec:model}. Different colors code the combination of the neurons which demonstrate spiking behavior for the given pair of values $(V_1, V_2)$. 

The plot distinguishes the following combinations of spiking neurons, denoted as:
\begin{itemize}
    \item $(0)$ -- none of the neurons are spiking;
    \item One-memristor spiking:
    \begin{itemize}
        \item $(1)$ -- $R_{M_1}$ spikes,
        \item $(2)$ -- $R_{M_2}$ spikes,
        \item $(3)$ -- $R_{M_3}$ spikes;
    \end{itemize}
    \item Two-memristor spiking:
    \begin{itemize}
        \item $(1,2)$ -- $R_{M_1}$ and $R_{M_2}$ spike,
        \item $(1,3)$ -- $R_{M_1}$ and $R_{M_3}$ spike,
        \item $(2,3)$ -- $R_{M_2}$ and $R_{M_3}$ spike;
    \end{itemize}
    \item Three-memristor spiking:
    \begin{itemize}
        \item $(1,2,3)$ -- all memristors ($R_{M_1}$, $R_{M_2}$, and $R_{M_3}$) spike.
    \end{itemize}
\end{itemize}
Remarkably, the above spiking patterns have distinctive areas on the parameter plane $(V_1,V_2)$, with larger areas corresponding to two- and three-memristor spiking regimes. 

At low voltages, none of the memristors exhibit spiking behaviour. However, as one of the voltages exceeds a threshold of approximately 145, a single-neuron spiking regime occurs in the system, with either $R_{M_1}$ or $R_{M_2}$ firing. Further increasing $V_{1,2}$ leads to the development of two- and three-memristor spiking regimes. Interestingly, for moderate voltage values around $V_{1}, V_{2}\approx 170$, the system demonstrates a single-memristor spiking dynamics, where only $R_{M_3}$ generates spiking. This regime occurs when the values of the voltages $V_{1,2}$ are not sufficient to trigger firing of either $R_{M_1}$ nor $R_{M_2}$, but their combined noisy output can have a DC component exceeding the threshold needed for generating spikes in $R_{M_3}$. Moreover, the presence of noise can lead to the appearance of repeating noise-induced spikes, even if a DC component is slightly below the threshold for self-sustained spiking. At a specific moment, when the combined input of DC voltage and random voltage fluctuations surpasses the spiking threshold, the memristor generates a spike. If this occurs frequently enough, the spiking can be detected within the measurement time.    
  
For a larger level of noise, $2k_{B}\eta_{i} = 10^{-4}$ in the memristive elements, the map of the regimes on the $(V_1,V_2)$ plane does not change qualitatively. However, the areas for two- and three-memristor spiking become considerably larger, while the region of the non-spiking regime shrinks -- see Fig. \ref{Fig:shakenpotential} (b). These changes can be explained by the fact that larger noise can produce observable noise-induced spiking for lower values of the applied DC voltage. This is illustrated in Fig. \ref{Fig:shakenpotential} (d)-(f) where bigger values of $(CV_1,CV_2)$ occur as noise-induced spiking dominates. For the case of even larger noise, $2k_{B}\eta_{i} = 10^{-3}$ as presented in Fig. \ref{Fig:shakenpotential} (c), the three-memristor spiking regime becomes dominant and can occur even if one of the voltages, $V_1$ or $V_2$, equals zero. 

Our study also shows that the area of existence of different patterns on the parameter plane $(V_1, V_2)$ depends on the values of circuit elements as well as the level of noise in the memristors. Note, that all the effects described above can be obtained even for a fixed $\beta$, and the fluctuating $\beta$ was used here for better match with experimental data below.

\subsection{Neuromorphic computing}
\label{ssec:NC}

The identified division of spiking regimes on the parameter plane $(V_{1}, V_{2})$ serves as a foundation for implementing synaptic convergence among three memristor blocks for brain-like computations \cite{Rolls:2020aa}, especially those involving signal comparison. 

Comparison operations are utilized in various brain computing functions, including {\it sensory discrimination}, {\it decision making}, {\it error detection}, and {\it pattern matching}. For instance, neurons in the prefrontal lateral cortex (LPFC) compare two sensory signals for pattern matching (discrimination) \cite{Wimmer9351} using inputs that have been processed by the sensory cortex and filtered by the thalamus to provide contextual relevance \cite{alonso2017thalamocortical}. Neurons in the hippocampus compare current input with stored input to emphasize differences between similar inputs or identify similarities to fill in missing information \cite{Olsen:2012aa}. Neuron cells in the cerebellum perform signal comparison for motor error correction \cite{Welnairz21}. 

In the case illustrated in Fig. \ref{Fig:shakenpotential} (a), the absence of spiking (pattern 0) indicates that both sensory voltages $V_{1,2}$ are small ($<140$). The appearance of pattern (1) suggests that $V_1\gg V_2$, but does not exceed 200. Pattern (2) indicates the opposite situation, where $V_2\gg V_1$, but does not exceed 200. Patterns (3) and (1,2,3) show that the sensory voltages $V_{1,2}$ are large and comparable, but in the former case do not exceed 210. 

The onset of pattern (1,3) tells that the sensory signal $V_1$ is much larger than $V_2$ if $V_2<100$, or comparable with $V_2$ if not. For the pattern (2,3), the situation is symmetrically opposite: $V_2 \gg V_1$ if $V_1 < 100$, or comparable with $V_2$ if not. 

Dividing the parameter plane $(V_1, V_2)$ into eight regions, each corresponding to one of the eight possible spiking patterns for a system of three neurons, facilitates a fundamental brain computation known as classification. Classification function enables categorization, the cognitive ability to group different objects as the same. In this context, each region on the $(V_1, V_2)$ plane can be assigned to a specific class of “objects”, which can be identified by analysing the features of the two classes represented by the values of $V_1$ and $V_2$. If different pairs of $V_1$ and $V_2$, representing different “objects”, induce the same spiking pattern, the “objects” are considered to belong to the same class. 

If one of the input voltages is however fixed, e.g., $V_1$, the artificial neural block becomes a comparator with one input, $V_2$. This comparator returns different spiking patterns depending on the range to which the value of $V_2$ belongs. For the case of Fig. \ref{Fig:shakenpotential} (a), the memristor blocks allows one to verify whether the value of $V_2 > 140$ using the only available output, $V_{M_2}(t)$. The absence of spiking indicates that $V_2 < 140$, while spiking evidences the opposite case. If two outputs, $V_{M_2}(t)$ and $V_{M_3}(t)$, are available, the classification of $V_2$ into three voltage ranges becomes feasible by analyzing the onsetting patterns: (0), (2), or (2,3). In the more noisy case depicted in Fig. \ref{Fig:shakenpotential} (c), a four-range classification of $V_2$ could be achieved if voltages from all memristors are available for analysis, by registering the patterns (0), (2), (2,3), and (1,2,3).
Note, the examples considered also illustrate that the memristive block in Fig.~\ref{Fig:neuromorphiccircuit} provides a device architecture with one or two inputs and 1-3 outputs, depending on the requested protocols and the specific type of comparison.

The artificial neuron block under consideration  can also function as logic gates. In this setup, the input is encoded by the levels of $V_1$ and $V_2$, while the output is indicated by the presence (“1”) or absence (“0”) of a specific spiking pattern. For example, in Fig. \ref{Fig:shakenpotential} (a), an input of “0” is assumed if $V_{1,2} \approx 0$, and “1” if $V_{1,2} \approx 200$, with the spiking (or lack thereof) in $R_{M_3}$ representing the output. A combination of $V_1\approx200$ (“1”) and $V_2\approx0$ (“0”) results in pattern (1), indicating no spiking in $R_{M_3}$ and setting the output to “0”. Similarly, inputs $V_1\approx0$ (“0”) and $V_2\approx200$ (“1”) trigger pattern (2) with no spiking in $R_{M_3}$. However, when both $V_{1,2}$ are 200 (“1”), pattern (3) occurs, causing $R_{M_3}$ to spike and setting the output to “1”. Under these assumptions, the circuit performs an “AND” operation (see Table~\ref{table:Logicalgates} left). 
Interestingly, with higher noise levels, as shown in Fig. \ref{Fig:shakenpotential} (c), the neuromorphic logic gate performs an “OR” operation (see Table~\ref{table:Logicalgates} right). If either $V_1$ or $V_2$ is set to 200, the system exhibits patterns (2,3), (1,3), or (1,2,3), all leading to spiking in $R_{M_3}$. Conversely, when $V_{1,2}\approx0$, no spiking occurs. Additionally, by creating a one-input block for Fig. \ref{Fig:shakenpotential} (a) with $V_2$ fixed around 200 and interpreting the output as the presence (“1”) or absence (“0”) of spiking in $R_{M_2}$, the circuit performs a “NOT” operation on the input $V_1$. This is because $V_1\approx0$ results in the pattern (2) with spiking in $R_{M_2}$, while $V_1\approx200$ changes the system to the pattern (3) where $R_{M_2}$ does not spike.

Although the Boolean operations (AND, OR, NOT, XOR etc.) do not form the core computation model in neuromorphic systems, they still play an important supporting role. For instance, memristor-based spiking neurons are designed with threshold-driven “all-or-nothing” logic and even dynamic logic like XOR using Boolean-like gating \cite{Duan:2020aa}.  Field Programmable Gate Array (FPGA) and CMOS implementations of spiking neurons and plasticity mechanisms rely on digital logic circuits (Boolean gates and finite state machines) to manage neuron states and timing \cite{Wang:2015aa}. Network-on-chip routers for spike communication use Boolean control signals and gating logic to multiplex and route spike events. Studies on spiking processors emphasize converting binary spike values into communication events and controlling them via Boolean logic \cite{Nazari:2023aa}. Finally, the Boolean gates are used in mixed-signal neuromorphic computing architectures \cite{Qiao2017}. Hence, the memristive synaptic convergence offers a promising bridge between neuromorphic analogue and digital architectures of computing systems.

\begin{figure*}[t!]
    \centering
    \includegraphics[width=6.85cm]{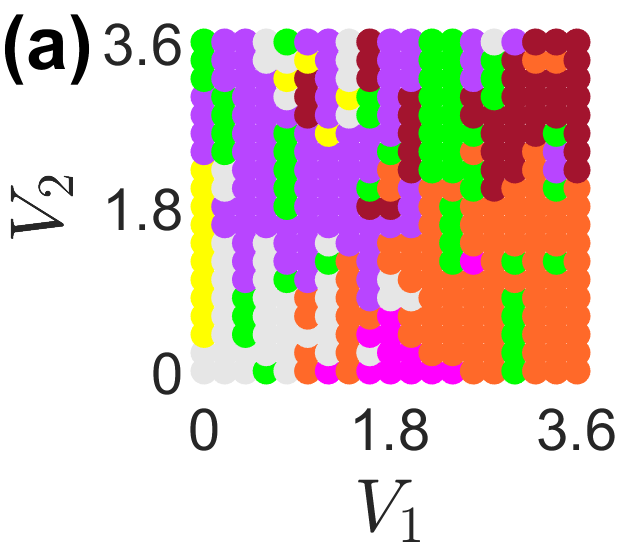}
    \includegraphics[width=6.85cm]{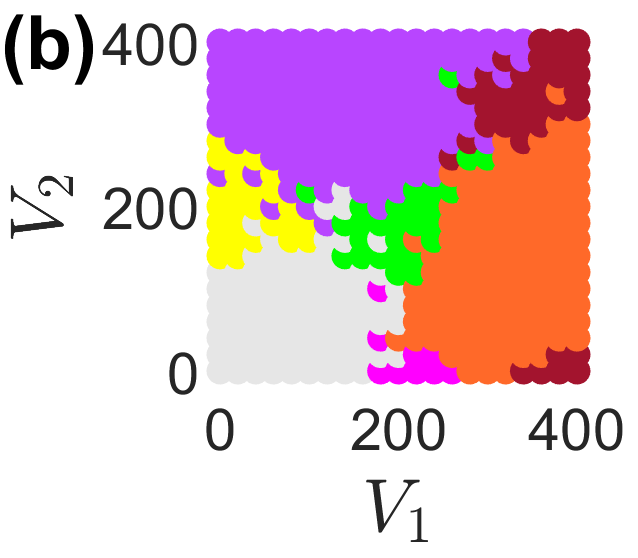}
    \caption{Regions of different spiking patterns measured experimentally (a) for two input voltages $V_1$ and $V_2$ in volts (V) applied to two artificial neurons with load resistances $R_L = 55$ k$\Omega$ and $60$ k$\Omega$, respectively. Similar spiking patterns are observed numerically (b) for $2k_{B}\eta_{1} = 10^{-4}$, $2k_{B}\eta_{2} = 10^{-3}$ and $2k_{B}\eta_{3} = 2.9 \times 10^{-4}$ with randomized $\beta$ and voltage step 20.}
    \label{Fig:theoryandexperimentalresults}
\end{figure*}

\begin{table}[htp!]
\begin{tabular}{|c|c|c|c|}
     \hline
     $R_{M_1}$ & $R_{M_2}$ & $R_{M_3}$ & Spiking  \\ \hline
     $0$ & $0$ & $0$ & $(0)$ \\ \hline
     $0$ & $1$ & $0$ & $(2)$ \\ \hline
     $1$ & $0$ & $0$ & $(1)$ \\ \hline
     $1$ & $1$ & $1$ & $(1,2,3)$\\ \hline
\end{tabular}
\hspace{0.2cm}
\begin{tabular}{|c|c|c|c|}
     \hline
     $R_{M_1}$ & $R_{M_2}$ & $R_{M_3}$ & Spiking  \\ \hline
     $0$ & $0$ & $0$ & $(0)$  \\ \hline
     $0$ & $1$ & $1$ & $(2,3)$ \\ \hline
     $1$ & $0$ & $1$ & $(1,3)$ \\ \hline
     $1$ & $1$ & $1$ & $(1,2,3)$ \\ \hline
\end{tabular}
\caption{Our three neuron circuit can realize an AND Gate (left) and OR GATE (right) for parameters $2k_{B}\eta_{i} = 10^{-5} $, $2k_{B}\eta_{i} = 10^{-4} $, and $2k_{B}\eta_{i} = 10^{-3}$ observed in Fig.~\ref{Fig:shakenpotential} depending on the chosen values of $V_1$ and $V_2$. For $V_1\approx0$ (“0”) and $V_2\approx200$ (“1”) trigger pattern (2) with no spiking in $R_{M_3}$ (see left: AND Gate) while $V_1\approx0$ (“0”) and $V_2\approx300$ (“1”) trigger pattern (2,3) with spiking in $R_{M_3}$ (see right: OR Gate) when $2k_{B}\eta_{i} = 10^{-5} $.}
\label{table:Logicalgates}
\end{table}
 
In summary, the memristive synaptic convergence depicted in Fig. \ref{Fig:neuromorphiccircuit} could function as a universal building block for implementing a range of brain-like computations related to input comparison. Additionally, it could be utilised in neuromorphic architectures for executing Boolean operations or for interfacing digital and neuromorphic elements and systems.

\subsection{Experimental measurements}

To verify the theoretical predictions,  we performed a series of experiments using the devices and methods discussed in sub-section \ref{ss:exp}. For the range of the voltages $V_{1,2} \in [0,3.6]$ V we registered the spiking patterns which are summarized in Fig.~\ref{Fig:theoryandexperimentalresults} (a). The colormap is identical to that in Fig. \ref{Fig:shakenpotential} and represents spiking patterns recorded over 15 seconds for a pair of $V_{1,2}$ with a voltage step of 0.2 V. 

The plot reveals that the regions of various spiking patterns are divided as predicted by the theoretical model \eqref{xirealization}-\eqref{Vm3realization}, thus enabling the fabricated device to perform the computational functions discussed in subsection \ref{ssec:NC} above. However, the measured spiking regions exhibit a more complex and erratic structure, where a large area corresponding to one pattern, e.g., $(1,3)$, includes small areas of other patterns, e.g., $(0)$, $(1)$, and $(3)$.

\begin{figure*}[t!]
    \centering
    \includegraphics[width=5.85cm]{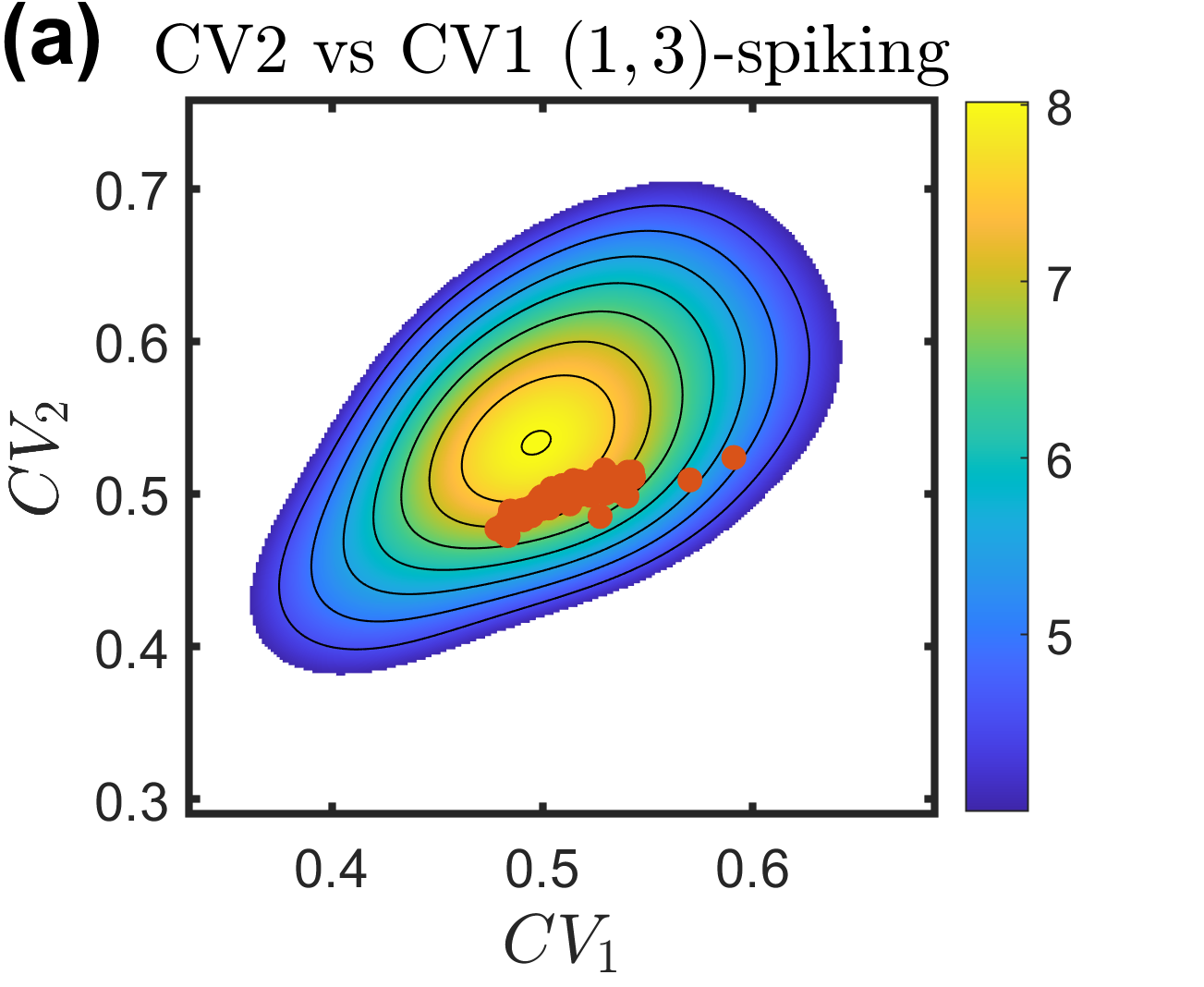}
    \includegraphics[width=5.85cm]{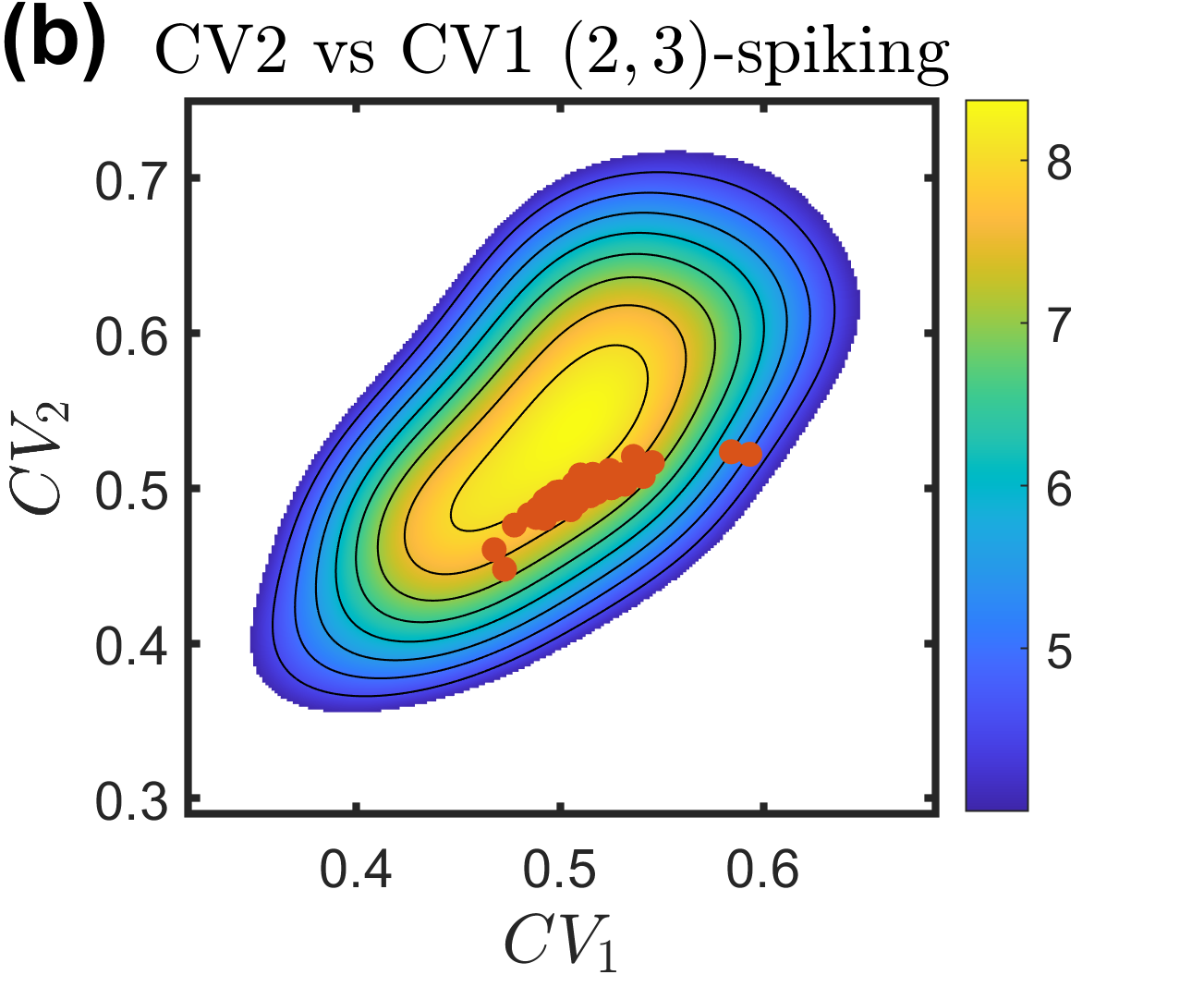}
    \includegraphics[width=5.85cm]{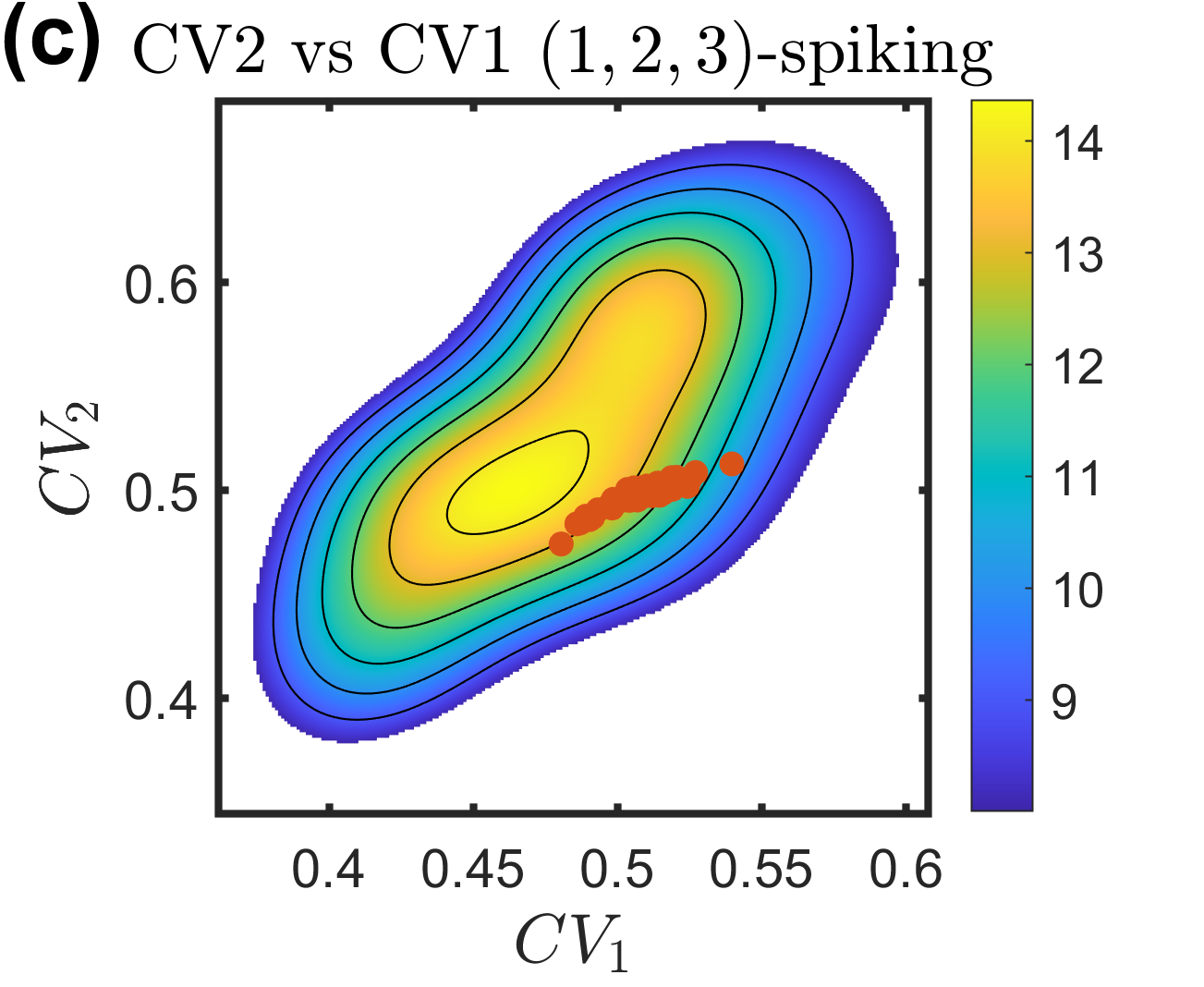}
    \caption{Joint probability density function for $CV_{1}$ and $CV_{2}$ calculated numerically (coloured contour plot) and scattered plot representing the pair of $CV_1$ and $CV_2$ calculated using the experimental data for the pattern $(1,3)$ (a), $(2,3)$ (b) and $(1,2,3)$ (c). The spikes were regist as detailed in \ref{ssec:model}.}
    \label{Fig:CV1CV2ExpTheory}
\end{figure*}
To comprehend the sporadic inclusion of other regions within the dominant pattern’s area of existence, we conducted several numerical simulations with varying levels of noise ($k_B\eta_{1,2,3}$) in the memristors. These simulations were performed with $\beta$ randomised as detailed in subsection \ref{ssec:model}. The results suggested that while small “islands” with patterns like $(1,2,3)$ could emerge in regions of simpler patterns such as $(1,3)$ and $(2,3)$ under the condition of $\beta=1$ [see Fig. \ref{Fig:shakenpotential} (b),(c)], the appearance of other inclusions, such as areas of $(3)$ within the $(2,3)$ pattern or $(0)$ within the $(1)$, $(2)$, or $(3)$ patterns, necessitates the randomization of $\beta$. The latter implies accounting for the random evolution of the filament \cite{savel2011molecular} in the description of particle dynamics. This finding is illustrated in Fig.~\ref{Fig:theoryandexperimentalresults} (b). The plot presents the results of calculations with randomized $\beta$, for $2k_{B}\eta_{1} = 10^{-4}$, $2k_{B}\eta_{2} = 10^{-3}$, and $2k_{B}\eta_{3} = 2.9 \times 10^{-4}$. The input voltages $V_{1,2}$ were varied within the range $[0,400]$ with a voltage step of $20$, while spiking was recorded over a time period of $t=5$. The resulting map of the regimes clearly shows the inclusion discussed above, qualitatively reproducing the results of the experiment in Fig.~\ref{Fig:theoryandexperimentalresults} (a).

To ensure that the theoretical model reproduces not only the characteristic patterns and their existence areas on the parameter plane $(V_1, V_2)$, but also the statistical properties of these patterns, we compare the global and local variability quantities $CV_1$ and $CV_2$, calculated using equations \eqref{eq:cv1} and \eqref{eq:cv2}, respectively. Fig.~\ref{Fig:CV1CV2ExpTheory} displays contour plots of the joint probability density function $p(CV_1, CV_2)$, numerically calculated for the theoretical model \eqref{xirealization}-\eqref{Vm3realization} using time realisations of $V_{M_3}(t)$ in cases where the patterns $(1,3)$ (a), $(2,3)$ (b), and $(1,2,3)$ (c) are realised in the block. For calculations, we set the parameter values as in Fig. \ref{Fig:theoryandexperimentalresults} (b), and for estimating $p(CV_1, CV_2)$, we used the function ``ksdensity” on Matlab R2022b. The figure demonstrates that the probability densities have different shape and slightly different positions of the maxima.
To compare the calculated $p(CV_1, CV_2)$ with the experimental data, we overlay the contour plots with scattered plots representing the experimentally measured pairs $(CV_1, CV_2)$ for each pattern. For all patterns, the clouds of experimentally measured points align well with the vicinities of the maxima of the probability density functions calculated numerically, indicating the good accuracy of our model.

Notably that the ranges of $CV_1$ and $CV_2$ ($0.35$ -- $0.7$) shown in Fig. \ref{Fig:CV1CV2ExpTheory} and corresponding to their most probable values fall well within the range of values typical for cortical neurons, which are  $0.2$  -- $1.5$  for both $CV_{1}$ and $CV_{2}$ \cite{ponce2010comparison,softky1993highly}. This fact highlights an interesting prospect of direct modeling the system of cortical neurons by the memristive device under study \cite{midya2025artificial}.

\section{Summary and Conclusion}

We theoretically and experimentally investigated the cooperative stochastic dynamics of a memristive circuit that simulates synaptic convergence among three spiking neurons. The circuit consists of two input block of artificial memristor-based neurons, which converge their outputs to the input of third neuron as shown in  Fig.~\ref{Fig:neuromorphiccircuit}.  As active elements in our artificial neurons we used diffusive memristors, where switching between resistive states is associated with formation and disruption of the conducting filament formed by metallic clusters diffusing in a dielectric matrix.  Specifically, we analysed how the presence or absence of spiking in each artificial neuron depends on the DC voltages $V_1$ and $V_2$ applied to the input neurons.  

Our theoretical model  \eqref{xirealization}-\eqref{Vm3realization} predicted that different combinations of $V_1$ and $V_2$ result in various dynamical patterns, where one, a mixture of two, or all three neurons are spiking. Remarkably, the voltage values associated with these spiking patterns create distinct areas on the parameter plane $(V_1, V_2)$, as shown in Fig~\ref{Fig:shakenpotential}. The presence of these distinct regions enables the circuit to be utilized in various neuromorphic computations related to comparing input signals and as logical elements performing Boolean operations, as discussed in subsection \ref{ssec:NC}.

The theoretical predictions align well with the experimental measurements. This is also confirmed by the agreement in the statistics of spikes, which has been quantified by the characteristics $CV_1$ and $CV_2$ calculated using equations \eqref{eq:cv1} and \eqref{eq:cv2}. However, the experimentally observed regions of existence for various spiking patterns exhibit a less uniform and more erratic structure compared to the initial theoretical results. Further numerical simulations suggest that this discrepancy may be due to the nonstationarity of the conducting filament, which continually changes shape and structure as electric current flows through the memristor. Incorporating this process into the model enhanced the agreement between theory and experiment (see Fig. \ref{Fig:theoryandexperimentalresults}). Therefore, stabilising the evolution of the conducting filament could reduce volatility and improve the reproducibility of the comparison functions provided by the memristive block. This stabilisation could be achieved through various methods, such as applying strict current compliance\cite{Carstens2021}, dedicated material and structure engineering \cite{Wang:2024aa}, or post-fabrication conditioning such as applying gamma radiation \cite{Pattnaik2023g}. Alternatively, different types of filament memristors, such as ovonic threshold switches \cite{Lee2020pra, Chai2020Wei}, could be used, offering more controlled and stationary conducting filaments.

Concurrently, the volatility in comparison functions due to the non-stationarity in filament formation in diffusion memristors presents a promising avenue to emulate the phenomenon known in neuroscience as neural variability. Neural variability refers to the natural fluctuations or differences in how neurons respond to the same stimulus over time or across trials. Even with identical inputs, a neuron’s output can differ. Recently, it has been suggested that this phenomenon can enhance brain computations \cite{Ponce2013, Festa:2021aa}. While experiments with living neurons are complex, it would be intriguing to explore the role of this phenomenon in neuromorphic computing \cite{Roadmap2024,gabayre2025advancements}. Specifically, it would be beneficial to investigate whether neural variability could lead to improved computations in the simplest systems of a few neurons, similar to those studied in this paper.

In conclusion, our findings provide a direction for developing universal computation blocks that could be used in the design of large-scale neuromorphic computing systems. They also offer a platform for better understanding the computing mechanisms and capabilities of living neural systems, particularly in cases where experiments with living cells are too complex or costly.

\section*{Acknowledgments}

The authors would wish to thank Dr Chris Mellor for his contributions to sample preparation.

\bibliography{diffmemreferences}

\end{document}